# Comparison of the band alignment of strained and strain-compensated GaInNAs QWs on GaAs and InP substrates

**Beşire Gönül *, Koray Köksal and Ebru Bakır**

Department of Engineering Physics, University of Gaziantep, 27310, Gaziantep, Turkey

## Abstract

We present a comparison of the band alignment of the $Ga_{1-x}In_xN_yAs_{1-y}$ active layers on GaAs and InP substrates in the case of conventionally strained and strained-compensated quantum wells. Our calculated results present that the band alignment of the tensiley strained $Ga_{1-x}In_xN_yAs_{1-y}$ quantum wells on InP substrates is better than than that of the compressively strained $Ga_{1-x}In_xN_yAs_{1-y}$ quantum wells on GaAs substrates and both substrates provide deeper conduction wells. Therefore, tensiley strained $Ga_{1-x}In_xN_yAs_{1-y}$ quantum wells with In concentrations of $x \leq 0.53$ on InP substrates can be used safely from the band alignment point of view when TM polarisation is required. Our calculated results also confirm that strain compensation can be used to balance the strain in the well material and it improve especially the band alignment of dilute nitride $Ga_{1-x}In_xN_yAs_{1-y}$ active layers on GaAs substrates. Our calculations enlighten the intrinsic superiority of N-based lasers and offer the conventionally strained and strain-compensated $Ga_{1-x}In_xN_yAs_{1-y}$ laser system on GaAs and InP substrates as ideal candidates for high temperature operation.

*Corresponding auther; bgonul@gantep.edu.tr,

Tel; 90 342 3601200/2214,   Fax; 90 342 3601100

Recently, there has been much interest in dilute nitride compound semiconductors of $Ga_{1-x}In_xN_yAs_{1-y}$ since the incorporation of nitrogen in $Ga_{1-x}In_xAs$ has a profound influence on the electronic properties of these materials and allows widely extended band structure engineering. Numerous experimental and theoretical work have been published about the $Ga_{1-x}In_xN_yAs_{1-y}$ on GaAs substrates [1-4]. For laser applications this material system has several important advantages as compared to the most commonly used GaInAsP/InP systems. First of all, a better high temperature performance of the laser structures is achieved due to a larger conduction band offset and, thus, improved electron confinement and decreased electron spill out at room temperature and above, thereby minimizing the threshold current in these lasers. Secondly, the increase of the electron effective mass with the addition of nitrogen provides a close match between the effective mass values for electrons and holes, beneficial for laser applications. Moreover, GaInNAs gives the flexibility of tailoring in the bandgap and an increase in the lattice parameter. Hence GaInNAs gives the potential to produce material lattice matched or mismatched to GaAs with a wide range of bandgap energies (from $\approx 1.5$ eV to less than 0.8 eV) [5].

Most attention to date has been focused on $Ga_{1-x}In_xN_yAs_{1-y}$ quantum wells (QWs) onGaAs substrates. Recently, some reports [6-8] demonstrated that strained and lattice-matched $Ga_{1-x}In_xN_yAs_{1-y}$ alloys on InP can also extend the wavelength of photonic device operation beyond that accessible to the $Ga_{1-x}In_xN_yAs_{1-y}$ /InP system. Increasing In content in $Ga_{1-x}In_xAs$ on InP beyond 53% results in a decrease of the band gap energy which, however, is partially offset by increasing compressive strain. Adding N to $Ga_{1-x}In_xAs$ reduces the band gap energy even further. Moreover, the incorporation of N compensates for compressive strain in the case of x>0.53 and introduces tensile strain in the case of $x \leq 0.53$, resulting an additional reduction of band gap energy for both cases. There are limited works so far on the growth of strained $Ga_{1-x}In_xN_yAs_{1-y}$ on InP substrates [6-14] and there is no work about the band alignment of the $Ga_{1-x}In_xN_yAs_{1-y}$ alloys on InP substrates up to our knowledge. In this work we will concentrate on the relative band alignment of the band edges of $Ga_{1-x}In_xN_yAs_{1-y}$ /InP between the quantum well and the barrier which is very important for modelling semiconductor quantum well structures. We will present the results in the case of tensiley strained $Ga_{1-x}In_xN_yAs_{1-y}$ quantum wells with In concentrations of $x \leq 0.53$ which provides the possibility of reaching TM mode emission at 1.55 $\mu m$ and above in telecommunication band which is difficult to reach with standart InGaAsP strained QW structures [14]. The GaInNAs/InP system combined with confinement layers, such as InAsP, could lead to the fabrication of tensile strained QW lasers as well as the development of optical isolators requiring TM polarisation [7, 8].

It has been shown previously that the band alignment of GaInNAs/GaAs can be analysed by means of model solid theory ignoring the presence of nitrogen in the average valence band energy values and taking into account the presence of nitrogen in all other laser parameters [15, 16]. The calculated results were in agreement with experimental data [17]. Therefore, we will use the model solid theory to calculate the band alignment of GaInNAs alloys on InP substrates as well. The composition dependence of band gap energy of GaInNAs will be determined by means of band anticrossing model [18]. This model is based on the interaction of the lowest conduction band with the highly localized N-induced energy level $E_N$, located 1.64 eV above the valence band edge of GaAs. We have used the band anticrossing model with an interaction parameter of $C_{MN}$=2.3 eV for tensiley strained

GaInNAs quantum wells on InP subtrates [13] and $C_{MN}$=2.7 eV for compressively strained GaInNAs quantum wells on GaAs subtrates [19].

Although, the incorporation of nitrogen in GaInNAs allows emission wavelengths as long as 1.55 $\mu m$ to be reached [20], the optical material quality deteriorate significantly with increasing N mole fractions [21], resulting in a much higher threshold current density of GaInNAs/GaAs lasers compared with that of GaInAs/GaAs lasers. In order to improve the performance of GaInNAs/GaAs quantum well lasers, the nitrogen composition of GaInNAs well should be reduced, although, this leads to an increased strain in the quantum wells. By introducing a strain-compensated barrier to this system it is possible to grow highly strained GaInNAs wells free of misfit dislocations. In strain-compensated QWs opposite strains are introduced in the well and barrier regions. These opposite strains balance each other and the average strain in the structure is reduced. In addition, for some laser configurations, such as short cavity lasers or distributed Bragg reflector lasers, a large number of QWs may be required for optimal performance [22]. As the number of strained QWs is increased, the total strain in the structure accumulates and the total strained layer thickness approaches a critical thickness at which lattice misfit dislocations start to form [23]. In strain-compensated QWs, the well width and the total number of wells can thus be increased, leading to an enhanced optical confinement. By means of introducing strains of opposite signs in the well and barrier layers to simultaneously vary the offsets of the heavy- and light-hole states, it is also possible to reduce the mixing between heavy- and light-hole states by means of spatially seperating them to different layers [24]. Experimental results [25, 26] showed that the strain-compensated QW lasers are desirable for optical applications with low threshold current and high efficiency. In addition, strain compensation gives access to a wider range of material composition, and thus improved possibilities to select band-edge offsets tailored to specific needs [27]. Therefore, there has been an interest in strain-compensated quantum well structures. This work investigates how the unique features of GaInNAs/GaAs and GaInNAs/InP quantum wells offer the best band alignment by means of using model solid theory. We also investigate the effect of the strain compensation on the band-alignments for the GaInNAs on GaAs and InP substrates. These calculations provide the first clear comparison of the strain-compensated band-alignment of the GaInNAs on two common substrates as a laser system and will enable us to predict the effect of the strain-compensation on band-offset energies for quantum well laser structures. For the band structures of laser systems, the material parameters except for the band gap energies are linearly interpolated from those of binary materials [28, 29]. We calculate the bulk band gap energy of GaInNAs by means of band-anti-crossing model [18] and the details of the calculations can be found in our recent paper [30].

We first present the band alignment of $Ga_{1-x}In_xN_yAs_{1-y}$ on GaAs substrates, which we consider to be a reference necessary for the understanding of the following results derived for low In containing $Ga_{1-x}In_xN_yAs_{1-y}$ on InP substrates. Figure 1 presents the calculated band offset ratios ($Q_c$ and $Q_v$ )and band offsets ($\Delta E_c$ and $\Delta E_v$ ) for uncompensated compressively strained $Ga_{0.70}In_{0.30}N_yAs_{1-y}$ quantum wells with GaAs barriers on GaAs substrates. The addition of N to GaInAs causes substantial changes in the band alignments; adding N to GaInAs increases the conduction band offset ratio $Q_c$ and decreases the valence band offset ratio $Q_v$. At first a rapid and then a gradual change in band offset ratios have been calculated. The corresponding band offsets are also shown in figure 1 which illustrates the fact that the addition of nitrogen into GaInAs leads the N-containing system having a band alignment of that of the ideal case (deep conduction- and shallow valence-wells) certainly. On the other hand, the addition of nitrogen deterioates laser parameters, so one should keep the N composition as low and low N leads high strains. So strain compensation can be offered as a solution in this novel material system and this can be achieved by means of using GaAsP barriers instead of GaAs barriers.

Figure 2 illustrates the effect of compensation of the compressive strain in the well by means of applying tensile strain in the barrier. The well is $Ga_{0.70}In_{0.30}N_yAs_{1-y}$ and the barrier is $Ga_{1-x}AsP_x$ and compressive strain in the well / tensile strain in the barrier is varied by means of increasing nitrogen N concentration in well / phosphorus P concentration in barrier, correspondingly, see figure 2. The uncompensated system corresponds to zero P concentration in the barrier. Upper curve represent the variation of the conduction band offset $\Delta E_c$ and lower curve represent the valence band offset $\Delta E_v$ with nitrogen concentration in well and phosphorus concentration in barrier, correspondingly, for $Ga_{0.70}In_{0.30}N_yAs_{1-y}$ quantum well. First, it should be noticed from the variations of figure 2 that $\Delta E_v$ (lower curve) decreases and $\Delta E_c$ (upper curve) increases with increasing N concentration. It should also to be noted from figure 2 that strain compensation by means of using GaAsP barriers instead of GaAs barriers improves the band alignment since $\Delta E_c$ further increases with an increase of P concentration in barrier. As an overall, by means of strain compensation, i.e. the increase of P concentration in barrier, it is possible to get deeper wells, leading to much beter confinement both in conduction and valence band. Therefore, strain compensation improves the band alignment and brings great advantageuos to this laser system of $Ga_{1-x}In_xN_yAs_{1-y}$ on GaAs substrates.

Strain compensation can also be achieved by means of using GaAsN barriers instead of GaAsP barriers. The band alingment of this strain compensated system is much better than that of the GaAsP barriers since the introduction of N into both the well and barrier yields deeper conduction wells (upper curve) and shallow valence wells (lower curve) as shown in figure 3. Therefore, the introduction of nitrogen into the barrier to compensate the compressive strain into the well of GaInNAs leads to this laser system on GaAs substrates having a band alignment of that of the ideal case.

We now want to compare the band alignment of $Ga_{1-x}In_xN_yAs_{1-y}$ QWs on GaAs substrates with that of the InP substrates. It is well known that the $Ga_{1-x}In_xAs$ can be grown lattice matched to InP for $x = 0.535$. Lowering the indium content or introducing N into the $Ga_{1-x}In_xAs$ induces a tensile strain. A sufficiently high tensile strain will lead the light-hole, lh, band being above that of the heavy-hole, hh, band. So fundamental transition will be due to $c_1$-$lh_1$ giving rise a TM-mode polarization emission. There are some reports [31-34] which emphasize the potential advantages of tensile strained QW lasers in terms of threshold current density, radiative characteristics, gain and loss mechanisms. Therefore, it is interesting to investigate the band alignment of $Ga_{1-x}In_xN_yAs_{1-y}$ quantum wells under tensile strain grown on an InP substrate for 1.5-1.6 $\mu m$ emission wavelength.

We first present the band alignment of conventionally tensiley strained $Ga_{1-x}In_xAs$ wells with unstrained InP barriers, as shown in figure 4, according to the model solid theory. As can be seen from figure 4, the valence band offset ratio $Q_v$ is greater than that of the conduction band offset ratio $Q_c$ which is not desirable for high temperature operation. It is seen from figure 4 that lowering In concentration of tensiley strained $Ga_{1-x}In_xAs$ alloy decreases both conduction and valence band offset at the expense of increasing the tensile strain in the well. In addition, the valence band offset $\Delta E_v$ is greater than the conduction band offset $\Delta E_c$. The right-hand-side of the y-axis of figure 4 shows the energy difference between the transition energies of $c_1$-$lh_1$ and $c_1$-$hh_1$ from the respective band edges. The negative energy difference shows that the polarization transition is TM since light-hole band lies above the heavy-hole band.

We want to illustrate at this point that the introduction of N to the well of $Ga_{1-x}In_xAs$ improves the band alignment on InP substrate significantly ; $\Delta E_c$ becomes greater than $\Delta E_v$ as shown in figure 5. The nitrogen concentration is held fixed at 2.5%. It is clear from figure 5 that $\Delta E_c$ increases rapidly and $\Delta E_v$ decreases with decreasing In concentration, i.e. by means of increasing tensile strain in the QW. The comparison of figure 5 with that of figures 1-3 reveals the fact that tensiley strained $Ga_{1-x}In_xN_yAs_{1-y}$ quantum wells on InP substrates have a deeper conduction well than that of the compressively strained $Ga_{1-x}In_xN_yAs_{1-y}$ quantum wells on GaAs substrates.

On the other hand, in order to have much deeper conduction wells than that of the valence band one should keep the In concentration as low as possible, and this leads to the problem of the critical thickness due to the high tensile strain in the well. Therefore we consider a strain compensated laser device of GaInNAs/InAsP/InP; the well composition is held fixed with a 10% In and 3% N and the arsenide concentration of the barrier of InAsP is varied. This results a compressive strain in the barrier to compensate the tensile strain in the well. Figure 6 illustrates the effect of the strain compensation on the band alignment for this laser system; the valence band offset ratio $Q_c$ is being greater than that of the conduction band offset ratio $Q_v$ for the stated range of As concentration and both $\Delta E_c$ and $\Delta E_v$ increases with increasing As concentration. It is seen from figure 6 that compensation decreases the $Q_c$ and increases the $Q_v$ value. Therefore the compensation can be thought as to bring disadvantageuos to this laser system. Fortunately, this is not the case. Although $Q_c$ and $Q_v$ shows opposite trend with compensation, both conduction- and valence-band offsets increases with compensation, as shown in figure 6. This strange behaviour of the increase of $Q_v$ with decreasing $Q_c$ can be explained as follows; the variation of the conduction band offset is a result of the combined effect of the varition of $Q_c$ and $\Delta E_g$ since $\Delta E_c = Q_c \Delta E_g$ and $\Delta E_g$ is the difference of the strained bandgap of the barrier and well. The rapid increase in $\Delta E_g (= E_{g_{barrier}} - E_{g_{well}})$ with As concentration eliminates the effect of the decrease in $Q_c$ with increasing As concentration and as an overall results an increase in $\Delta E_c$ with increasing As concentration. So compensation again can be considered a solution to balance the tensile strain in the well of GaInNAs alloy on InP substrates as in the case GaAs substrates.

As a summary, the design of GaInNAs based devices requires a deep knowledge of the alloy's electronic properties and a development of accurate models. Band alignment of the QWs is one of the important properties while modelling the QW structures since this property determines the high temperature performance of the laser structures. Therefore, this work presents theoretical calculations to compare the band alignments of the N-free and N-included laser devices on GaAs and InP substrates emitting in the neighborhood of 1.3 and 1.6 μm. Our calculations indicate that the band alignment of the N-based conventionally strained QW laser systems on InP is better than that of the GaAs substrate and both substrates provide deeper conduction wells. Therefore, tensiley strained $Ga_{1-x}In_xN_yAs_{1-y}$ quantum wells with In concentrations of $x \leq 0.53$ on InP substrates can be used safely from the band alignment point of view when TM polarisation is required. The introduction of opposite strain to the barrier in N-based lasers on both GaAs and InP substrates not only result deep electron wells but also cause the electron wells being much deeper than that of the hole wells which is essential to have good high temperature characteristics. Therefore, these calculations enlighten the intrinsic superiority of N-based lasers and offer conventionally strained the strain-compensated N-based laser system on GaAs and InP substrates as ideal candidates for high temperature operation.

## Figure Captions

Figure 1 The nitrogen N concentration (in well) dependence of the conduction and valence band offset ratios, $Q_c$ and $Q_v$, (inset figure) and the corresponding conduction and valence band offsets, $\Delta E_c$ and $\Delta E_v$, of the uncompensated compressively strained $Ga_{0.70}In_{0.30}N_yAs_{1-y}$ quantum wells with GaAs barriers on GaAs substrates.

Figure 2 The variation of conduction (upper curve) and valence (lower curve) band offsets, $\Delta E_c$ and $\Delta E_v$, with nitrogen concentration which results compressive strain (in well) for $Ga_{0.70}In_{0.30}N_yAs_{1-y}$ quantum wells and with phosphorus concentration which results tensile strain (in barrier) for compensated $Ga_{0.70}In_{0.30}N_yAs_{1-y}$ / GaAsP quantum wells on GaAs substrates.

Figure 3 The calculated variation of the conduction (upper curve) and valence (lower curve) band offsets, $\Delta E_c$ and $\Delta E_v$, with nitrogen concentration which results compressive strain (in well) for $Ga_{0.70}In_{0.30}N_yAs_{1-y}$ quantum wells with GaAs barriers on GaAs substrates and with nitrogen concentration which results tensile strain (in barrier) for compensated $Ga_{0.70}In_{0.30}N_{0.025}As_{0.975}$ /GaAsN quantum wells on GaAs substrates.

Figure 4 The indium concentration (in well) dependence of the conduction and valence band offset ratios, $Q_c$ and $Q_v$, (inset figure) and the corresponding the conduction and valence band offsets, $\Delta E_c$ and $\Delta E_v$, of the tensiley strained uncompensated $In_xGa_{1-x}As$ / InP QW laser system on InP substrates. RHS of the y-axis shows the energy difference between the transition energies of $c_1$- $lh_1$ and $c_1$- $hh_1$ from the respective band edges.

Figure 5 The variation of the conduction and valence band offsets, $\Delta E_c$ and $\Delta E_v$, with indium concentration and tensile strain (in well) for uncompensated $Ga_{1-x}In_xN_{0.02}As_{0.98}$ /InP QWs on InP substrates. RHS of the y-axis shows the energy difference between the transition energies of $c_1$- $lh_1$ and $c_1$- $hh_1$ from the respective band edges.

Figure 6 The calculated variation of the conduction and valence band offsets, $\Delta E_c$ and $\Delta E_v$, with arsenide concentration and compressive strain (in barrier) for compensated $Ga_{0.90}In_{0.10}N_{0.03}As_{0.97}$ / $In_xAs_{1-x}P$ / InP laser system.

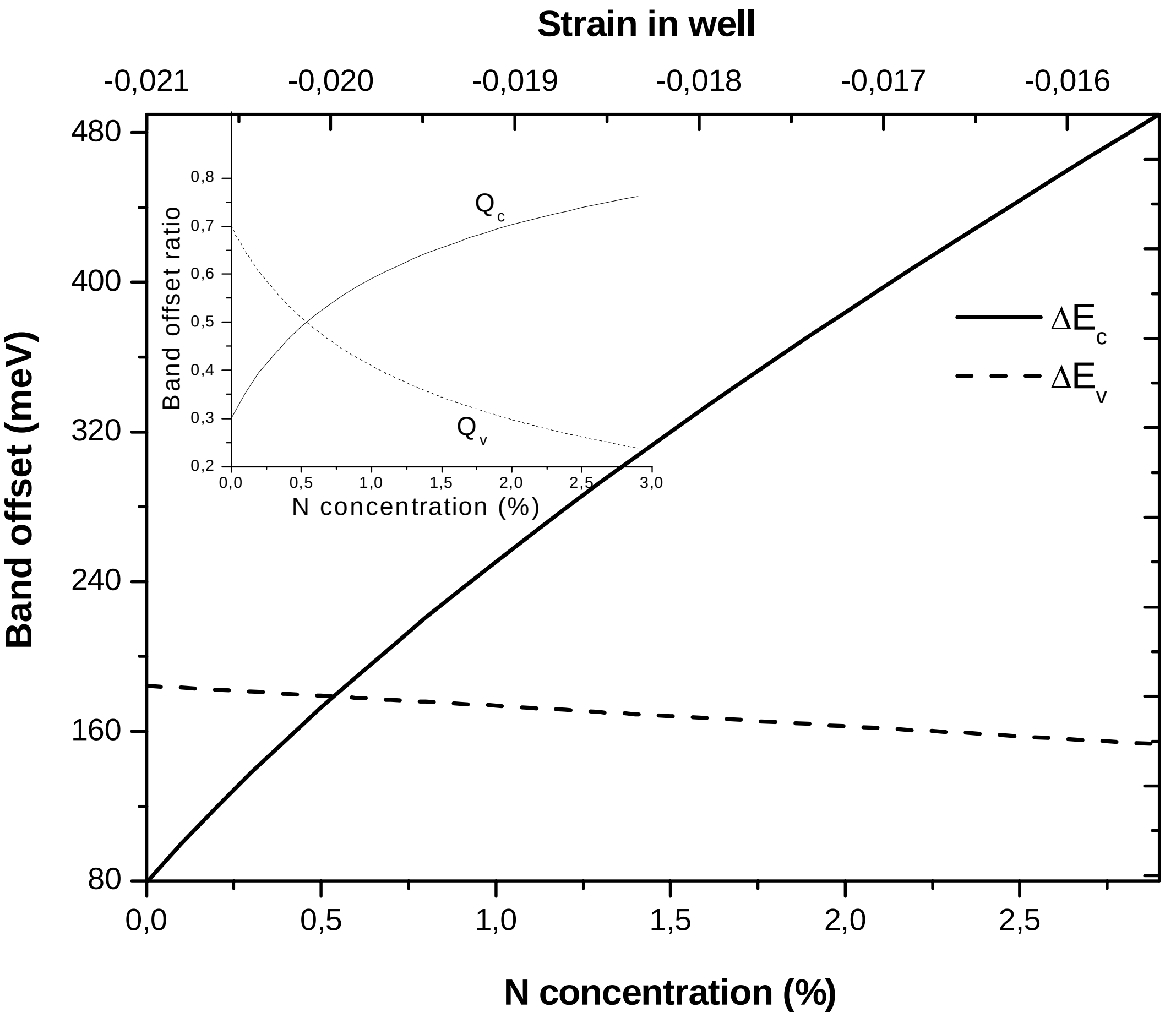

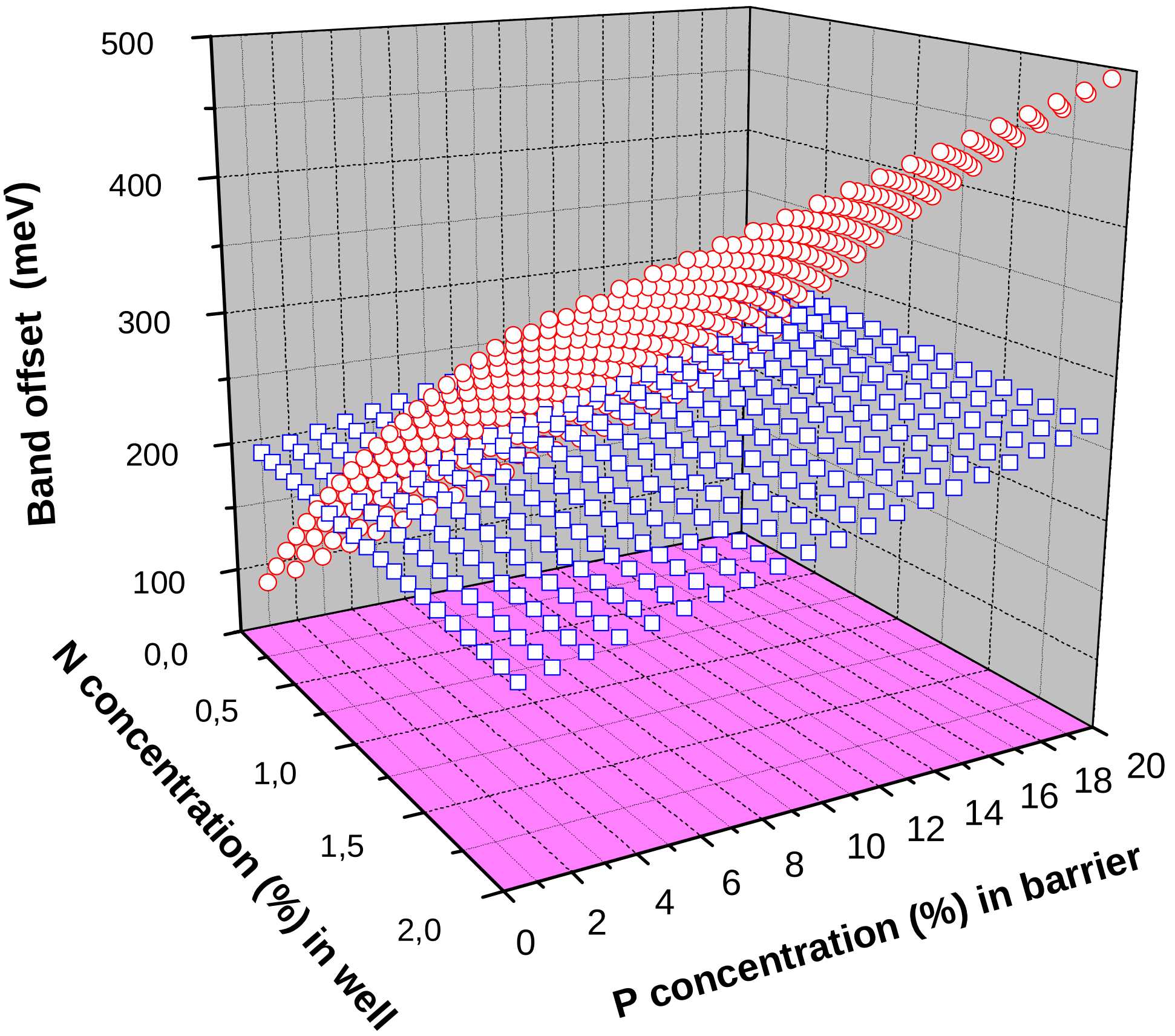

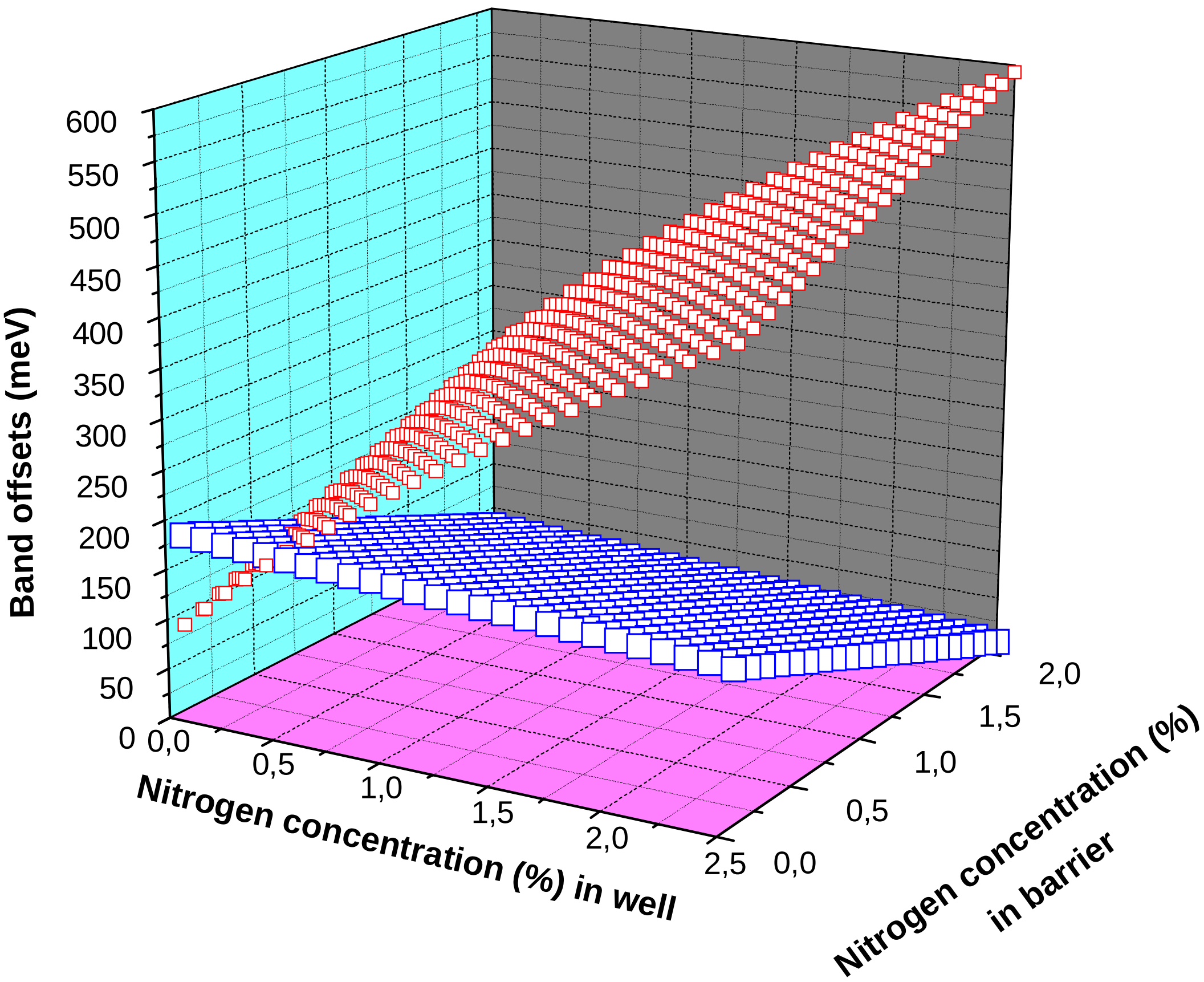

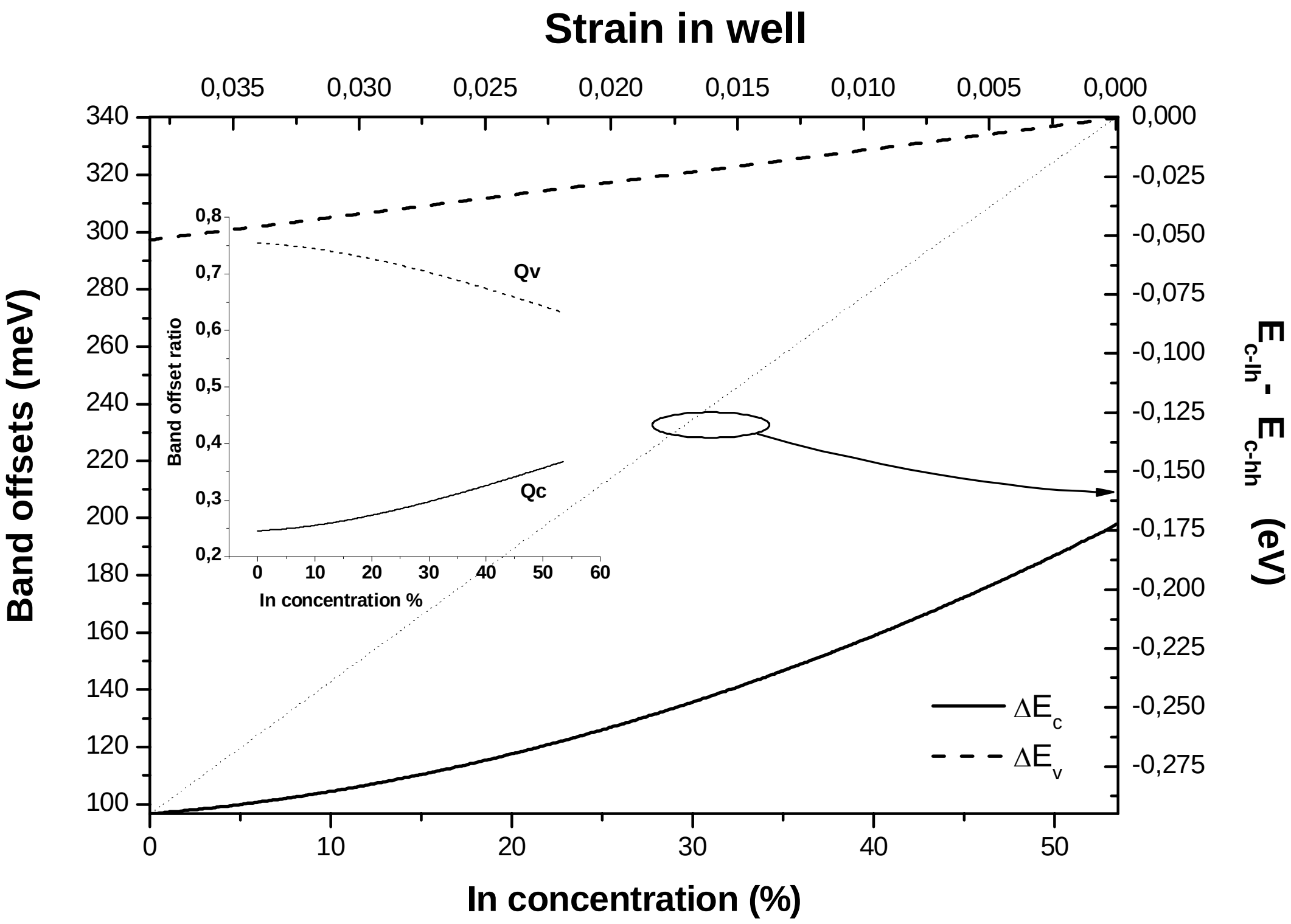

Strain in well
0,035 0,030 0,025 0,020 0,015 0,010 0,005 0,000
Band offsets (meV)
340 320 300 280 260 240 220 200 180 160 140 120 100
0,000 -0,025 -0,050 -0,075 -0,100 -0,125 -0,150 -0,175 -0,200 -0,225 -0,250 -0,275
E<sub>c-lh</sub> - E<sub>c-hh</sub> (eV)
0 10 20 30 40 50
In concentration (%)
ΔEc
ΔEv
Band offset ratio
0,8 0,7 0,6 0,5 0,4 0,3 0,2
Qv
Qc
0 10 20 30 40 50 60
In concentration %

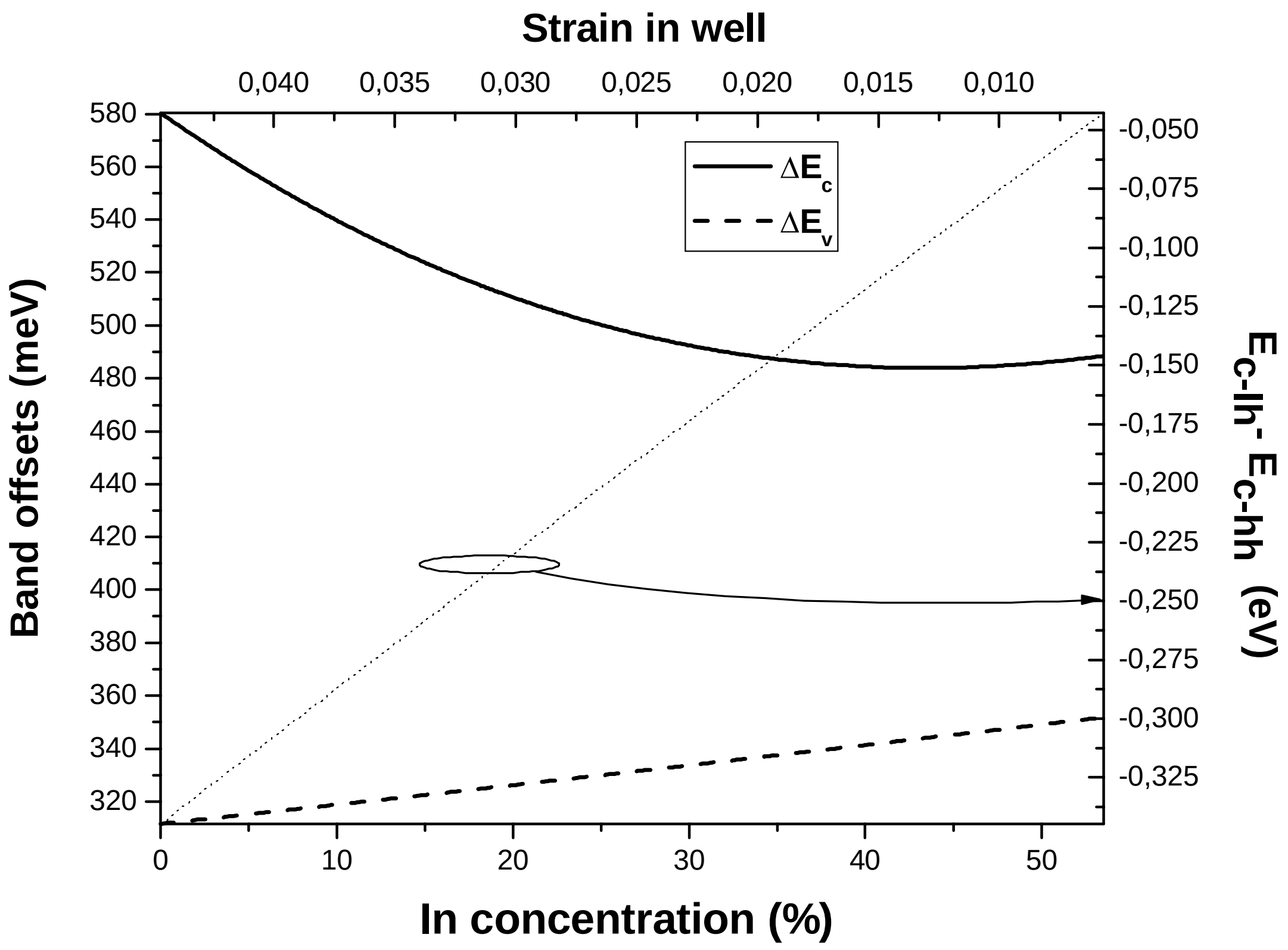

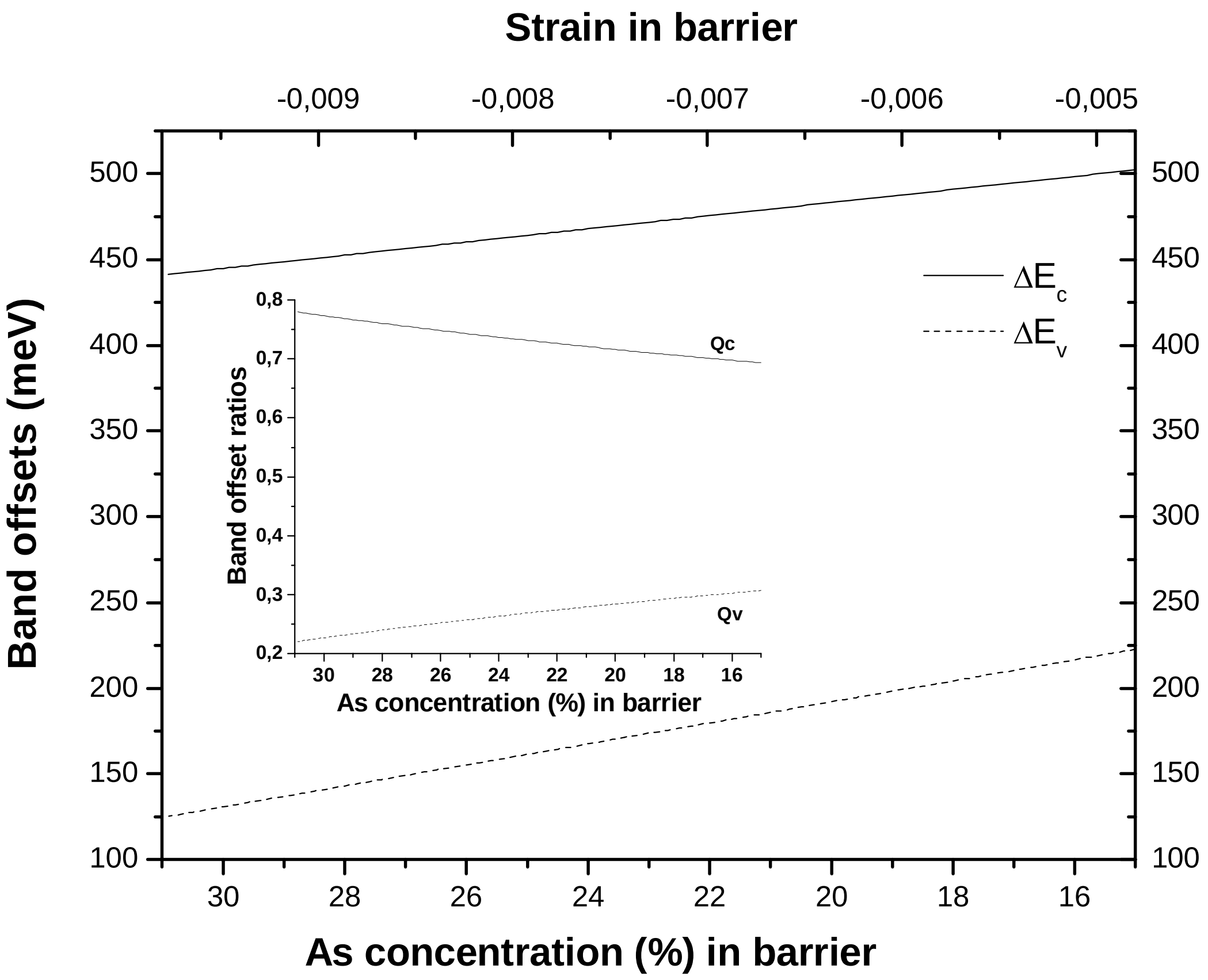